\RequirePackage[pagewise,mathlines]{lineno}

\documentclass[twocolumn,aps,showpacs,superscriptaddress]{revtex4-1}
\usepackage{savesym}
\usepackage{amssymb}
\usepackage{xcolor}
\usepackage{bbding}
\usepackage{graphicx}

\usepackage{dcolumn}
\restoresymbol{TXT}{affil}
\usepackage{bm}

\begin{document}
\pagewiselinenumbers

\def\Journal#1#2#3#4{{#1} {\bf #2}, #3 (#4)}

\def\NCA{Nuovo Cimento}
\def\NIM{Nucl. Instr. Meth.}
\def\NIMA{{Nucl. Instr. Meth.} A}
\def\NPB{{Nucl. Phys.} B}
\def\NPA{{Nucl. Phys.} A}
\def\PLB{{Phys. Lett.}  B}
\def\PRL{Phys. Rev. Lett.}
\def\PRC{{Phys. Rev.} C}
\def\PRD{{Phys. Rev.} D}
\def\ZPC{{Z. Phys.} C}
\def\JPG{{J. Phys.} G}
\def\CPC{Comput. Phys. Commun.}
\def\EPJ{{Eur. Phys. J.} C}
\def\PR{Phys. Rept.}
\def\PRV{Phys. Rev.}
\def\JHEP{JHEP}

\preprint{}
\title{Systematic study of the experimental measurements on ratios of  different $\Upsilon$ states}
\date{\today}
\author{Wangmei Zha}\email{wangmei@rcf.rhic.bnl.gov}\address{University of Science and Technology of China, Hefei, China}\address{Brookhaven National Laboratory, New York, USA}
\author{Chi Yang}\address{University of Science and Technology of China, Hefei, China}\address{Brookhaven National Laboratory, New York, USA}
\author{Bingchu Huang}\address{Brookhaven National Laboratory, New York, USA}
\author{Lijuan Ruan}\address{Brookhaven National Laboratory, New York, USA}
\author{Shuai Yang}\address{University of Science and Technology of China, Hefei, China}\address{Brookhaven National Laboratory, New York, USA}
\author{Zebo Tang}\address{University of Science and Technology of China, Hefei, China}
\author{Zhangbu Xu}\address{Brookhaven National Laboratory, New York, USA}
\begin{abstract}
The world data on yields and ratios of different $\Upsilon$ states created in hadron collisions at $\sqrt{s} = 19-8000$ GeV are examined in systematic way.  We find that $\Upsilon(2S)/\Upsilon(1S)=0.275\pm0.005$ and $\Upsilon(3S)/\Upsilon(1S)=0.128\pm0.004$.
No signficant energy dependence of these ratios are observed within the broad collision energies. In addition, the rapidity, transverse momentum, and transverse mass dependence of these ratios are also reported.

\end{abstract}
\maketitle

\section{Introduction}
Current data at RHIC demonstrate rapidly thermalizing matter characterized by: 1) initial energy densities far above the critical values predicted by lattice QCD for formation of a Quark Gluon Plasma (QGP); 2) opacity to jets; and 3) nearly perfect fluid flow, which is marked by constituent interactions of very short mean free path, established most probably at a stage preceding hadron formation ~\cite{starwhitepaper}. One of the important next objectives is to study the color screening length in QGP ~\cite{rhicIIQuarkonia}. Constituent heavy quark-antiquark pairs in quarkonia are subject to color screening and should be dissolved at high temperature. The dissociation temperatures of different species will vary due to different radii and binding energies. The precise measurements of transverse momentum ($p_T$) distributions of quarkonia for different centralities, collision systems, and energies will serve as a thermometer of the QGP.

$J/\psi$ is the most studied quarkonium at the RHIC and LHC. In central collisions, $J/\psi$  is less suppressed in Pb+Pb at 2.76 TeV for $2.5 \!< |y| \!< 4$ than in Au+Au collisions at 200 GeV for $1.2 < |y| < 2.2$ ~\cite{phenixjpsi,ALICEjpsi}. A similar feature is observed for mid-rapidity comparisons between LHC and RHIC energies. Comparisons to model calculations indicate that recombination plays a significant role at RHIC and a more significant role at LHC. Furthermore, the $p_T$ dependence of $J/\psi$ suppression ~\cite{ALICEjpsi, phenixjpsi:07, starjpsi:12}, the flow pattern of $J/\psi$ as a function of $p_T$ ~\cite{Yang:12, starjpsiv2}, and the high $p_T$ $J/\psi$ suppression pattern as a function of centrality ~\cite{starjpsi:12,Moon:12}, reported at LHC and RHIC, can be described consistently by model calculations incorporating color screening and recombination features (See a recent review article for details~\cite{GaleRuan:12}).

Besides $J/\psi$, the different $\Upsilon$ states are also ideal tools for this study since the ground state and excited states melt at different temperatures and all of them could decay to dileptons. Furthermore, since the   $b\bar{b}$ cross section at RHIC energy is expected to be two orders of magnitude smaller than $c\bar{c}$ cross section from FONLL calculations ~\cite{vogtum}, the recombination contribution from QGP phase might be negligible to bottomonium production, at least at RHIC. Meanwhile, the $\Upsilon$ absorption cross sections with the abundantly produced hadrons in these collisons are relatively small according to model calculation ~\cite{nuclear absorption}, therefore, suppression from the absorption effect is expected to be unimportant. These make $\Upsilon$ an even better probe for studying the color screening effect in QGP if sufficient statistics can be achieved experimentally. 

In addition to its important role in studying the color screening effect, a measurement of the $\Upsilon(1S, 2S, 3S)$ states in $p+p$ and heavy ion collisons can help to set limits on the medium temperature created in heavy-ion collisions, because the sequential suppression pattern of the excited states is sensitive to the temperature reached in the medium ~\cite{sequential suppression}. The nuclear modification factor $R_{AA}$, which can quantify the suppression of particle production in nuclear collisons, is defined as the ratio of the inclusive hadron yield in nuclear collisions to that in $p+p$ collisions scaled by the underlying number of binary nucleon-nucleon collisions.
 Measurements at CMS show that $R_{AA}$[$\Upsilon$(1S)] is 0.4-0.5 while $R_{AA}$[$\Upsilon$(2S)] is 0.1 in central Pb+Pb collisions at 2.76 TeV ~\cite{cmsupsilon:12}. At RHIC 200 GeV, $R_{AA}$[$\Upsilon$(1S+2S+3S)] is about 0.4 in central Au+Au collisions ~\cite{Trzeciak:12}. Considering feeddown contributions [~50\% for $\Upsilon$(1S) ], measurements at RHIC and LHC indicate that $\Upsilon$(3S) is completely melted and that $\Upsilon$(2S)  is strongly suppressed in central A+A collisions. 

With bremsstrahlung radiation of electrons traversing the detector material, the STAR experiment is at the borderline of being able to separate the three states. The muons in $\Upsilon$  do not suffer from this degradation. The Muon Telescope Detector (MTD) at STAR to be completed in 2013 will provide excellent muon identification capabilities and thus enable the measurements of different $\Upsilon$ states~\cite{mtd} through $\Upsilon \rightarrow \mu^+\mu^-$ to estimate the medium temperature created at RHIC.  In the RHIC-II era, with the full MTD system at STAR, we will be able to measure different $\Upsilon$ states with good precision by taking advantage of the high luminosity in Au+Au collisions. While in 200 GeV $p+p$ collisions which can provide the reference data for $R_{AA}$ calculation, the $\Upsilon(3S)$ total cross section measurement with 10\% statistical uncertainty on total cross section requires sample 400 $pb^{-1}$ luminosity, which may not be allowed by the run plan. On the other hand, different $\Upsilon$ states have been measured from low to high energy and we may use the previous measurements to derive the baseline of their ratios for the corresponding heavy ion collisions. In this article, we study the world-wide data and obtain different $\Upsilon$ state ratios versus energy.

\section{Results}
Measurements of the $\Upsilon$ production have been performed by many experiments ~\cite{J. Badier:NA3,J.K. Yoh,W.R. Innes,K. Ueno,S. Childress,G. Moreno:E605,T. Yoshida:E605,P.L. McGaughey:E772,T. Alexopoulos:E771,L. Camilleri:R-108,C. Kourkoumelis,A.L.S. Angelis:CCOR,C. Albajar:UA1,F. Abe:CDF,Upsilon:CMS,Upsilon:LHCb,Upsilon:ATLAS,Upsilon:E886,Upsilon:LHCb8TeV} in the wide range of the proton-nucleon center-of-mass energy $\sqrt{s}$ of 19 to 8000 GeV, and with targets ranging from proton (A = 1) to platinum (A = 195) with both proton and antiproton beams. Only few experiments could allow precise measurements of the dilepton mass spectrum in the range of 8-15 GeV/$c^{2}$ and seperation of the three lowest-mass $\Upsilon$ states ~\cite{J.K. Yoh,K. Ueno,T. Yoshida:E605,G. Moreno:E605,Upsilon:E886,F. Abe:CDF,Upsilon:CMS,Upsilon:LHCb,Upsilon:ATLAS,Upsilon:LHCb8TeV}. We explore the $p_T$, rapidity, and transverse mass ($m_T$) dependence of $\Upsilon(2S)$ and $\Upsilon(3S)$ production relative to the $\Upsilon(1S)$ by deriving the ratios with n=2,3, defined as following: \begin{equation}
\label{eq1}
 \frac{\Upsilon(nS)}{\Upsilon(1S)} = \frac{\sigma[p(\bar{p})+p(A) \rightarrow \Upsilon(nS)] \times Br[\Upsilon(nS) \rightarrow l^{+}l^{-}]}{\sigma[p(\bar{p})+p(A) \rightarrow \Upsilon(1S)] \times Br[\Upsilon(1S) \rightarrow l^{+}l^{-}]}
\end{equation} where $\sigma[p(\bar{p})+p(A) \rightarrow \Upsilon(nS)]$ (n=1,2,3) is the $\Upsilon(nS)$  cross section in $p(\bar{p})+p(A)$, $Br[\Upsilon(nS) \rightarrow l^{+}l^{-}]$ is the dilepton branching ratio of the corresponding $\Upsilon$ states. Such observables are sensitive to the magnitude and kinematic dependencies of feed-down contributions between the three $\Upsilon$ states. Production yields of quarkonium states could be modified from $p(\bar{p})+p$ to $p(\bar{p})+A$ collisions by cold nuclear matter effects ~\cite{R.Vogt}. However, such effects should have a small impact on the ratios. The nuclear modifications of the parton distribution functions should have an equivalent effect on the three $\Upsilon$ states, because partons involved into the production have similar kinematics. Nuclear dependence of $\Upsilon$ production measured at E772 ~\cite{Upsilon:E772} and E886 ~\cite{Upsilon:E886} indicates that no nuclear dependence difference,within uncertainties, is observed between the $\Upsilon(1S)$ and the sum of $\Upsilon(2S + 3S)$.

\renewcommand{\floatpagefraction}{0.75}
\begin{table*}[htbp]
\newcommand{\tabincell}
\centering
\begin{center}
\begin{tabular}{|c|c|c|c|c|c|c|c|}
\hline
experiment & system & energy (GeV) & rapidity & $\Upsilon(2S)/\Upsilon(1S)$ & $\Upsilon(3S)/\Upsilon(1S)$ & $\Upsilon(2S+3S)/\Upsilon(1S)$& ref. \\
\hline
CFS \textcolor{blue}{$\triangledown$}& $p+p$ & $19.4$ & $<y>_{ac} = 0.40$ & $0.670 \pm 0.940$ & $0.100 \pm 0.600$ &$0.770 \pm 1.115$& ~\cite{J.K. Yoh} \\
\hline
CFS \textcolor{blue}{$\triangledown$}& $p+p$ & $23.7$ & $<y>_{ac} = 0.21$ & $0.460 \pm 0.130$ & $0.000 \pm 0.080$ &$0.460 \pm 0.157$& ~\cite{J.K. Yoh} \\
\hline
CFS \textcolor{blue}{$\triangledown$}& $p+p$ & $27.4$ & $<y>_{ac} = 0.03$ & $0.380 \pm 0.110$ & $0.080 \pm 0.060$ &$0.460 \pm 0.122$& ~\cite{J.K. Yoh} \\
\hline
CFS \textcolor{blue}{$\vartriangle$}& $p+P_{t}$ & $27.4$ & $y = 0$ & $0.310 \pm 0.030$ & $0.150 \pm 0.020$ &$0.460 \pm 0.034$& ~\cite{K. Ueno} \\
\hline
E605  \textcolor{red}{$\blacktriangle$}& $p+B_{e}$ & $38.8$ & $y = 0$ & $0.310 \pm 0.110$ & $0.090 \pm 0.060$ &$0.400 \pm 0.125$& ~\cite{T. Yoshida:E605} \\
\hline
E605 \textcolor{red}{$\lozenge$}& $p+C_{u}$ & $38.8$ & $-0.15\!<x_{F} \!< 0.25$ ($-0.28\!< y \!< 0.46$) & $0.270 \pm 0.011$ & $0.131 \pm 0.008$ &$0.400 \pm 0.014$& ~\cite{G. Moreno:E605} \\
\hline
E886 \textcolor{red}{$\square$}& $p+d$ & $38.8$ & $0\!<x_{F} \!< 0.6$ ($0.00\!<y \!< 0.98$) & $0.321 \pm 0.012$ & $0.127 \pm 0.009$ &$0.448 \pm 0.016$& ~\cite{Upsilon:E886} \\
\hline
E886 \scriptsize{\textcolor{red}{$\bigcirc$}}& $p+p$  & $38.8$ & $0\!<x_{F} \!< 0.6$ ($0.00\!<y \!< 0.98$) & $0.274 \pm 0.017$ & $0.134 \pm 0.013$ &$0.408 \pm 0.022$& ~\cite{Upsilon:E886} \\
\hline
CDF \textcolor{green}{$\blacktriangledown$}& $p+\bar{p}$ & $1800$ & $|y| \!< 0.4$ & $0.281 \pm 0.048$ & $0.155 \pm 0.032$ &$0.436 \pm 0.058$& ~\cite{F. Abe:CDF} \\
\hline
CMS \large{\textcolor{black}{$\bullet$}} & $p+p$ & $7000$ & $|y| \!< 2.0$ & $0.258 \pm 0.012$ & $0.138 \pm 0.010$ &$0.396 \pm 0.015$& ~\cite{Upsilon:CMS} \\
\hline
LHCb \textcolor{black}{$\blacklozenge$} & $p+p$ & $7000$ & $2.0 \!< y \!< 4.0$ & $0.245 \pm 0.015$ & $0.124 \pm 0.008$ &$0.369 \pm 0.020$& ~\cite{Upsilon:LHCb} \\
\hline
ATLAS \textcolor{black}{$\blacksquare$}& $p+p$ & $7000$ & $|y| \!< 2.25$ & $0.256 \pm 0.019$ & $0.115 \pm 0.010$ &$0.371 \pm 0.024$& ~\cite{Upsilon:ATLAS} \\
\hline
LHCb \textcolor{black}{$\bigstar$}& $p+p$ & $8000$ & $2.0 \!< y \!< 4.5$ & $0.256 \pm 0.005$ & $0.125 \pm 0.003$ &$0.381 \pm 0.006$& ~\cite{Upsilon:LHCb8TeV} \\
\hline
\end{tabular}
\end{center}
\caption{(color online) $\Upsilon(2S)/\Upsilon(1S)$, $\Upsilon(3S)/\Upsilon(1S)$ and $\Upsilon(2S+3S)/\Upsilon(1S)$ in different experiments.}
\label{table1}
\end{table*}

\renewcommand{\floatpagefraction}{0.75}
\begin{figure}[htbp]
\begin{center}
\includegraphics[keepaspectratio,width=0.45\textwidth]{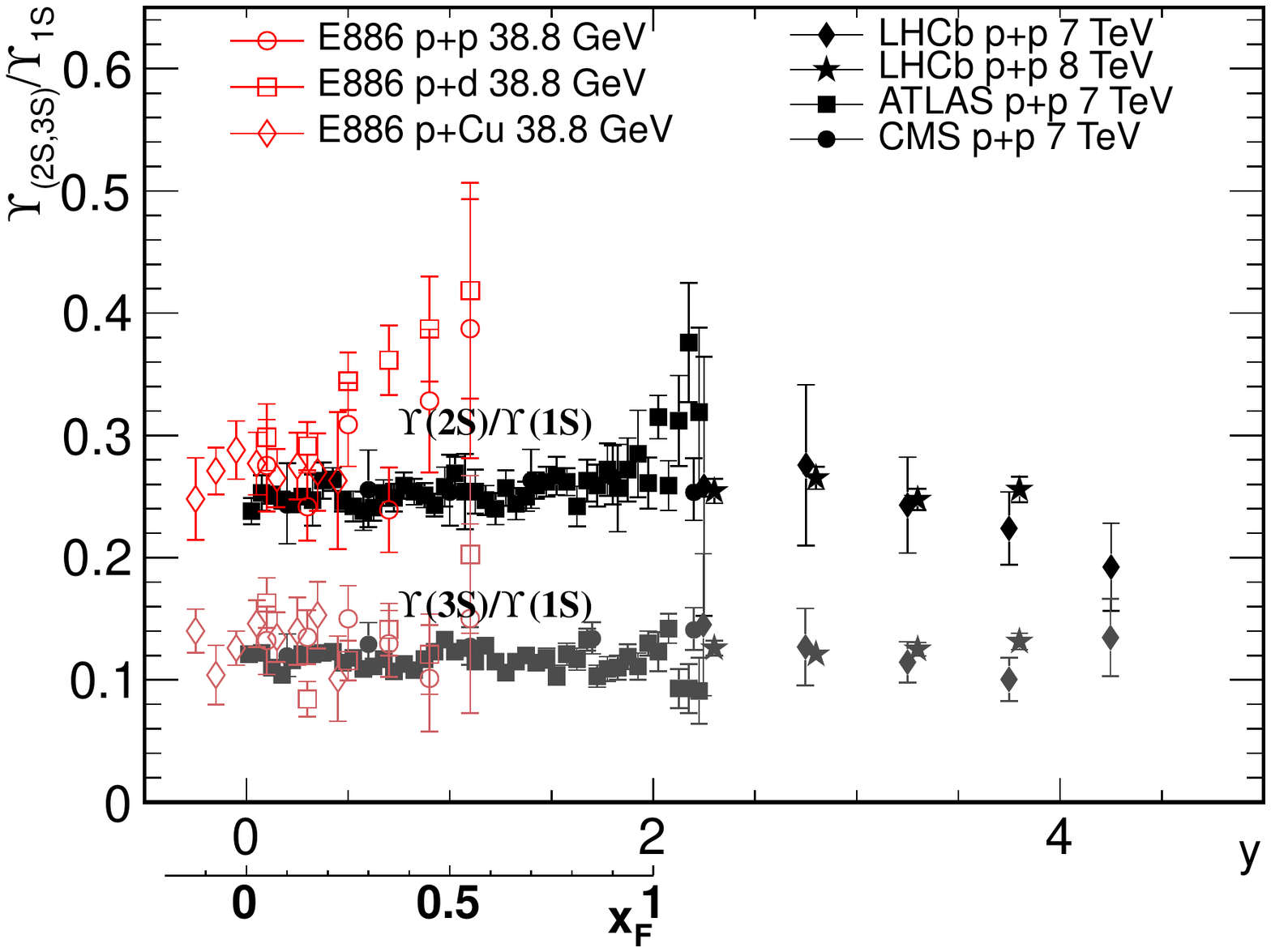}
\caption{(color online) Ratios of $\Upsilon(2S)/\Upsilon(1S)$ and $\Upsilon(3S)/\Upsilon(1S)$ as a function of rapidity (or feyman scale) measured by E886, ATLAS and LHCb experiments. The error bars represent the quadrature sum of statistical and systematic uncertainties.} 
\label{figure3}
\end{center}
\end{figure}
 Figure~\ref{figure3} shows the ratios of $\Upsilon(2S)/\Upsilon(1S)$ and $\Upsilon(3S)/\Upsilon(1S)$ as a function of $\Upsilon$ rapidity (or Feynman scaling variable $x_F$).The LHCb 7 TeV ~\cite{Upsilon:LHCb}, 8 TeV ~\cite{Upsilon:LHCb8TeV} and ATALS ~\cite{Upsilon:ATLAS} results are measured as a function of rapidity. The universal rapdity independent ratios for both $\Upsilon(3S)/\Upsilon(1S)$ and $\Upsilon(2S)/\Upsilon(1S)$ are observed from these three experiments  within their rapidity coverage. For fixed target experiments ~\cite{Upsilon:E886,G. Moreno:E605}, the ratios are extracted as a function of $x_F$. Taking the uncertainties into account, we can not claim any $x_F$ dependence of the ratios for both $\Upsilon(2S)/\Upsilon(1S)$ and $\Upsilon(3S)/\Upsilon(1S)$. These ratio values are almost the same as those measured at LHC despite the huge center-of-mass energy difference. 

\renewcommand{\floatpagefraction}{0.75}
\begin{figure}[htbp]
\begin{center}
\includegraphics[keepaspectratio,width=0.45\textwidth]{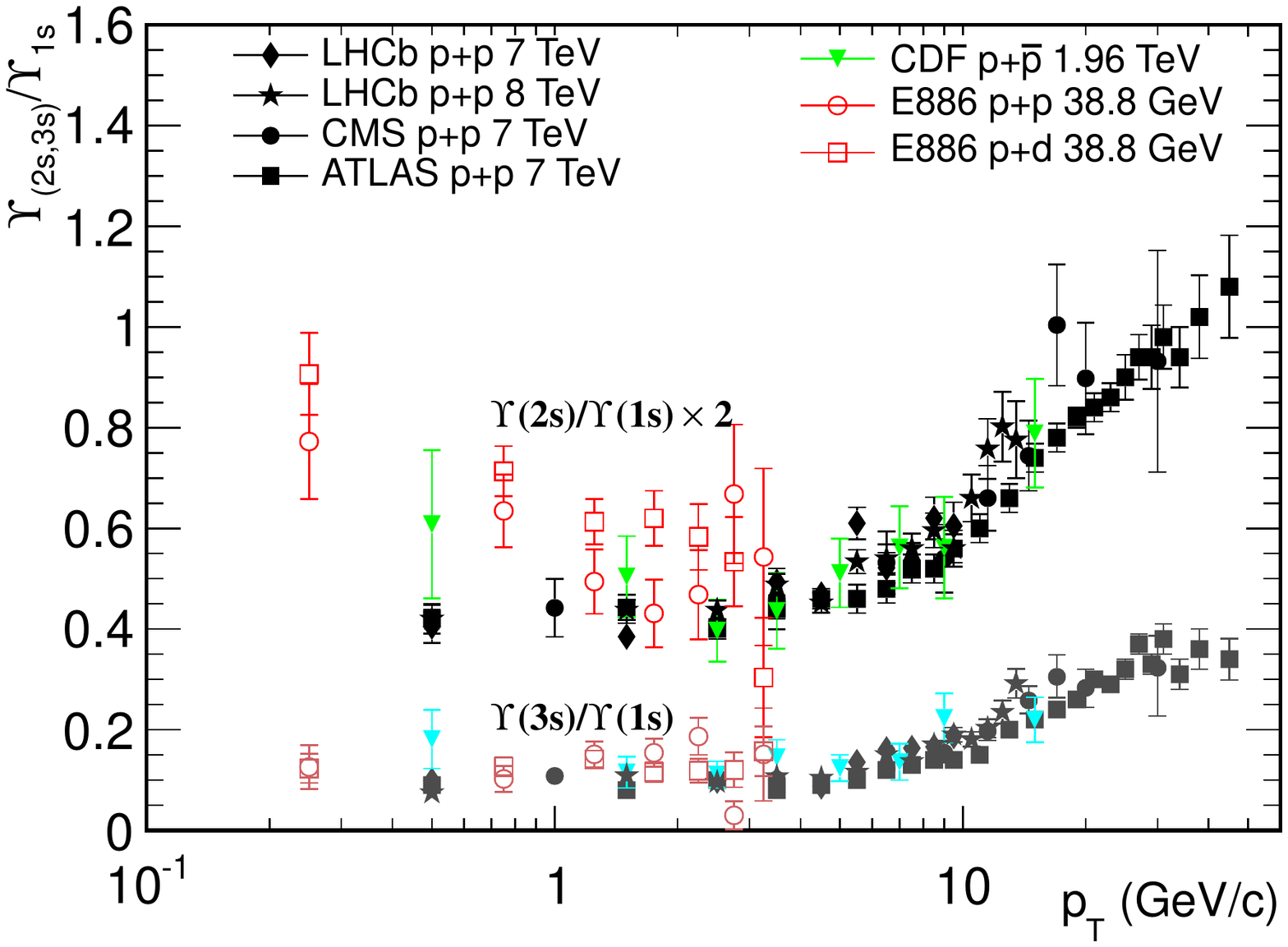}
\caption{(color online) Ratios of differential $\Upsilon(2S)/\Upsilon(1S)$ and $\Upsilon(3S)/\Upsilon(1S)$ as a function of transverse momentum measured in world-wide experiments.The error bars represent the quadrature sum of statistical and systematic uncertainties.} 
\label{figure2}
\end{center}
\end{figure}
Figure~\ref{figure2} depicts the ratios of $\Upsilon(2S)/\Upsilon(1S)$ and $\Upsilon(3S)/\Upsilon(1S)$ as a function of $p_{T}$. The measured $\Upsilon(2S,3S)/\Upsilon(1S)$ ratios at ATLAS ~\cite{Upsilon:ATLAS} are relatively constant in the $0\!<p_{T}\!< 5$ GeV/$c$ range. At higher $p_T$ a significant and steady rise in the relative production rates of higher $\Upsilon$ states is observed. The measured results at CMS ~\cite{Upsilon:CMS} and LHCb ~\cite{Upsilon:LHCb,Upsilon:LHCb8TeV} follow the same trend  as that of ATLAS for the corresponding $p_T$ ranges. The fixed target measurements ~\cite{Upsilon:E886} seem different in their $p_{T}$ coverage ($0\!<p_{T}\!<3.5$ GeV/$c$) for $\Upsilon(2S)/\Upsilon(1S)$, which decreases with $p_{T}$. 

\renewcommand{\floatpagefraction}{0.75}
\begin{figure}[htbp]
\begin{center}
\includegraphics[keepaspectratio,width=0.45\textwidth]{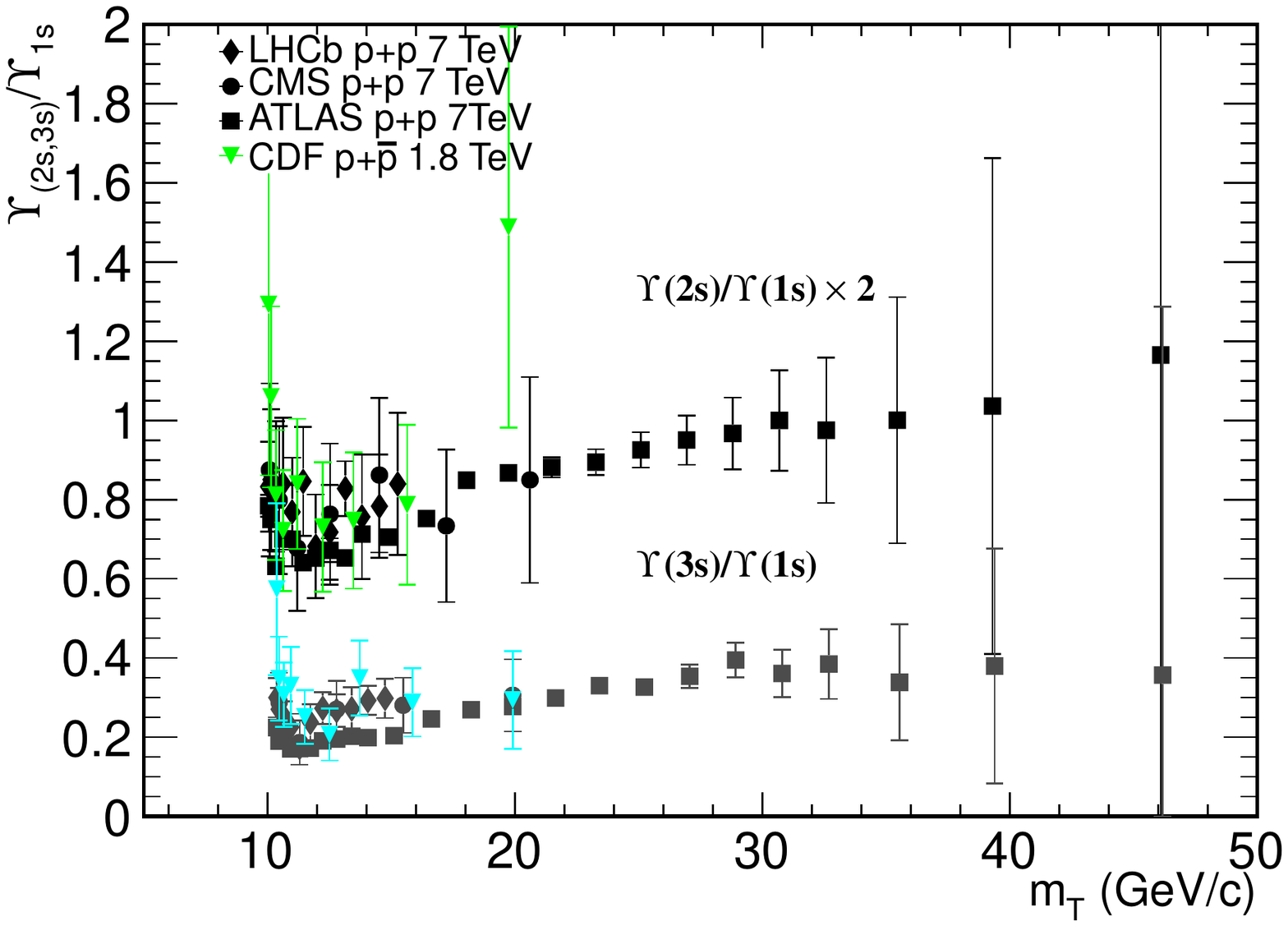}
\caption{(color online) Ratios of $\Upsilon(2S)/\Upsilon(1S)$ and $\Upsilon(3S)/\Upsilon(1S)$ as a function of $m_{T}$ measured in world-wide experiments. The error bars represent the quadrature sum of statistical and systematic uncertainties. 
} \label{figure1}
\end{center}
\end{figure}
Figure~\ref{figure1} shows the ratios of $\Upsilon(2S)/\Upsilon(1S)$ and $\Upsilon(3S)/\Upsilon(1S)$ as a function of $m_{T}$. The results at LHCb ~\cite{Upsilon:LHCb,Upsilon:LHCb8TeV}, CMS ~\cite{Upsilon:CMS}, ATLAS~\cite{Upsilon:ATLAS}, and CDF ~\cite{F. Abe:CDF} all follow the same trend. The ratios first decrease and then increase with the increasing $m_{T}$.

\renewcommand{\floatpagefraction}{0.75}
\begin{figure}[htbp]
\begin{center}
\includegraphics[keepaspectratio,width=0.45\textwidth]{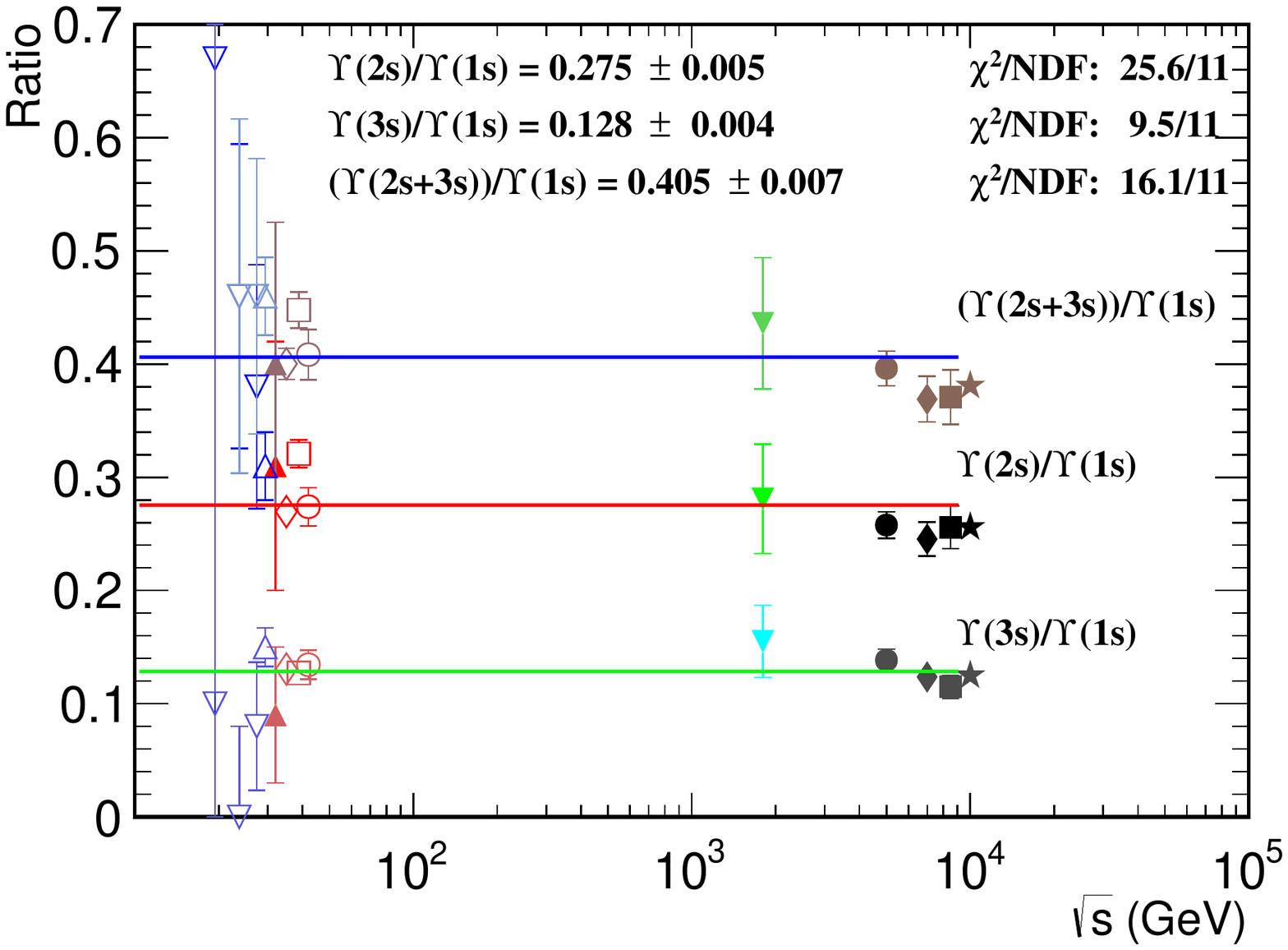}
\caption{(color online) Ratios of $\Upsilon(2S)/\Upsilon(1S)$, $\Upsilon(3S)/\Upsilon(1S)$ and $\Upsilon(2S+3S)/\Upsilon(1S)$ as a function of center-of-mass energy measured in world-wide experiments.The error bars represent the quadrature sum of statistical and systematic uncertainties.} 
\label{figure0}
\end{center}
\end{figure}
Figure~\ref{figure0} and Table~\ref{table1} show the $\Upsilon(2S)/\Upsilon(1S)$, $\Upsilon(3S)/\Upsilon(1S)$, and $\Upsilon(3S + 2S)/\Upsilon(1S)$ as a fuction of center-of-mass energy using the experimental measurements of the three $\Upsilon$ states ~\cite{J.K. Yoh,K. Ueno,T. Yoshida:E605,G. Moreno:E605,Upsilon:E886,F. Abe:CDF,Upsilon:CMS,Upsilon:LHCb,Upsilon:ATLAS}. The ratios with the same center-of-mass energy have been shifted slightly for clarity in the figure. As shown in Table~\ref{table1}, the center-of-mass energy points are for 19.4 GeV, 23.7 GeV, 27.4 GeV, 38.8 GeV, 1.8 TeV, and 7 TeV. The energy points with relatively precise ratios are 38.8 GeV, 1.8 TeV, and 7 TeV. The ratios measured at these three energy points are almost the same and show no energy dependence despite the huge center-of-mass energy difference. The constant fit results are  $\Upsilon(2S)/\Upsilon(1S)=0.275\pm0.005$ and $\Upsilon(3S)/\Upsilon(1S)=0.128\pm0.004$. This provides much more precise reference ratios in 200 GeV $p+p$ collisions.

\section{Summary}
We study the world-wide data of different $\Upsilon$ states at  $\sqrt{s} = 19-8000$ GeV. We find that $\Upsilon(2S)/\Upsilon(1S)=0.275\pm0.005$ and $\Upsilon(3S)/\Upsilon(1S)=0.128\pm0.004$.
No signficant energy dependence of these ratios are observed within the broad collision energies. In addtion, we observed no significant rapidity dependence of these ratios  within the broad collision energies. The $m_T$ dependence of these ratios measured in LHC and CDF follow the same trend.  The $p_T$ trend of the $\Upsilon(2S)/\Upsilon(1S)$ ratios seem different from each other in the region $0\!<p_{T}\!<3.5$ GeV/$c$ between LHC results and fixed target results. 

\section{Acknowledgments}
We express our gratitude to the STAR Collaboration and the RCF at BNL for their support. This work was supported in part by the U.S. DOE Office of Science under the contract No. DE-AC02-98CH10886; authors Wangmei Zha and Chi Yang are supported in part by the National Natural Science Foundation of China under Grant Nos 11005103 and 11005104.

\end{document}